\documentclass[lettersize,journal]{IEEEtran}
\usepackage{amsmath,amsfonts}
\usepackage{algorithmic}
\usepackage{algorithm}
\usepackage{array}
\usepackage{textcomp}
\usepackage{stfloats}
\usepackage{url}
\usepackage{verbatim}
\usepackage{graphicx}
\usepackage{caption}
\usepackage{hyperref}
\usepackage{balance}
\usepackage{subcaption}
\usepackage{cite}
\newcommand{\pol}[0]{\pmb{\pi}}
\newcommand{\cpol}[0]{\pmb{\mu}}

\usepackage{pifont}
\usepackage{xcolor}
\newcommand{\cmark}{\textcolor{green}{\ding{51}}}%
\newcommand{\xmark}{\textcolor{red}{\ding{55}}}%

\begin{document}

\title{LLM-based Multi-Agent Reinforcement Learning: Current and Future Directions}

\author{Chuanneng Sun,~\IEEEmembership{Student Member,~IEEE}, Songjun Huang,~\IEEEmembership{Student Member,~IEEE},\\
and Dario Pompili,~\IEEEmembership{Fellow,~IEEE}
\thanks{The authors are with the Department of Electrical and Computer Engineering, Rutgers University--New Brunswick, NJ, USA. Emails: \emph{\{chuanneng.sun, songjun.huang, pompili\}@rutgers.edu}}
\thanks{
This work was supported by the NSF RTML Award No. CCF-1937403.}
}

\markboth{Submitted to IEEE Robotics \& Automation Letters, May 2024}%
{Sun \MakeLowercase{\textit{et al.}}: LLM-based Multi-Agent Reinforcement Learning: Current and Future Directions}


\maketitle
\thispagestyle{empty}

\begin{abstract}
In recent years, Large Language Models~(LLMs) have shown great abilities in various tasks, including question answering, arithmetic problem solving, and poem writing, among others. Although research on LLM-as-an-agent has shown that LLM can be applied to Reinforcement Learning~(RL) and achieve decent results, the extension of LLM-based RL to Multi-Agent System~(MAS) is not trivial, as many aspects, such as coordination and communication between agents, are not considered in the RL frameworks of a single agent. 
To inspire more research on LLM-based MARL, in this letter, we survey the existing LLM-based single-agent and multi-agent RL frameworks and provide potential research directions for future research. In particular, we focus on the cooperative tasks of multiple agents with a common goal and communication among them. We also consider human-in/on-the-loop scenarios enabled by the language component in the framework.
\end{abstract}

\section{Introduction}

\IEEEPARstart{M}{ulti}-Agent Reinforcement Learning (MARL) has emerged as a popular approach to address the coordination problem in Multi-Agent Systems~(MAS). As opposed to Individual Reinforcement Learning~(IRL)-based or traditional optimization-based solutions, MARL has shown a significant improvement in scalability and robustness to uncertainty and dynamicity~\cite{sun2023hmaac, shalev2016safe, sadhu2020aerial, calvo2018heterogeneous}.
This improvement is largely attributed to the communication and coordination among agents inherent in MARL, where multiple agents learn and adapt their policies simultaneously while interacting within a shared environment and communicating with others.
However, how and what to communicate among the agents in the MAS remains to be explored. Representative examples include MARL frameworks that learn to generate numerical messages using neural networks, formulate neural communication protocols, and learn targeted ad hoc communications. Despite the decent performance of the MARL frameworks achieved in various applications, they still underperform human experts. As a result, it is reasonable to think \emph{why not leveraging human knowledge and human languages in MARL?}

\begin{figure}
    \centering
    \includegraphics[width=0.5\textwidth]{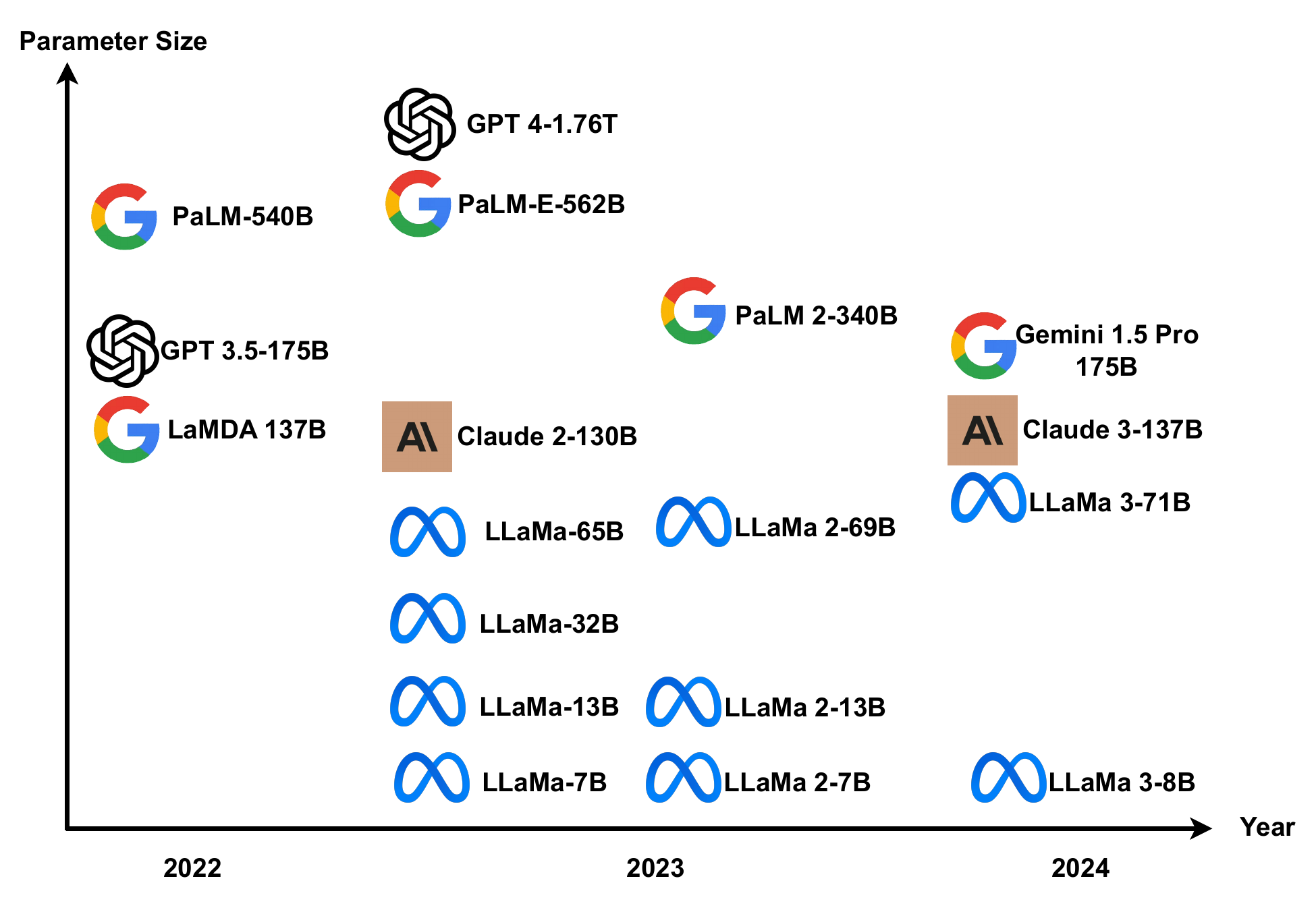}
    \caption{Well-known Large Language Models~(LLMs) over the past three years. Among them, only PaLM-E from Google is trained specifically for embodied applications, e.g., robot control.}
    \label{fig:llm}
\end{figure}

As recent advances in Natural Language Processing~(NLP) demonstrate great abilities in multi-modal tasks, language-conditioned MARL becomes a promising research problem. NLP has been an active research topic for decades and many famous models have been proposed for language modeling such as Recurrent Neural Network~(RNN)~\cite{rumelhart1986learning, jordan1997serial}, Long-Short Term Memory networks~(LSTM)~\cite{hochreiter1997long}, and transformers~\cite{vaswani2017attention}.
These foundational models have greatly improved the ability of machines to understand and generate human language, setting the stage for more complex applications.

In recent years, the integration of NLP with single-agent RL has led to the development of language-conditioned RL frameworks~\cite{peng2023conceptual, jiang2019language, zhou2021inverse}, especially as Large Language Models~(LLMs)~\cite{openai2023chatgpt, touvron2023llama, chowdhery2023palm, team2023gemini} emerged as the rising star in the artificial intelligence community (see Fig.~\ref{fig:llm}) and has been successfully applied in various fields~\cite{wu2024new, lai2024language, han2024chainofinteraction}. Pre-trained LLMs contain general human knowledge about the world and can easily adapt to RL problems without the need for retraining.
This integration not only leverages the semantic richness of language
but also allows for the dynamic adjustment of agent behaviors based on linguistic input. In particular, LLM is able to generate new information that it has not seen before on the basis of a few examples. For example, in Reflexion~\cite{shinn2024reflexion}, the authors showed that the LLM agent could generate decent reflections on its decisions without any reward/feedback from the environment. Such capabilities are particularly valuable in multi-agent systems, where agents must coordinate and cooperate based on shared goals communicated through language. 


Due to the need for communication and coordination, the problem of MARL becomes more complex than simply multiplying the RL of a single agent by the number of agents. As opposed to conventional MARL, LLMs-based MARL can leverage linguistic cues to facilitate inter-agent communication and collaboration, further boosting system performance. For example, agents can use shared language to negotiate roles, coordinate actions, or exchange information about the environment or their internal states, thereby aligning their objectives more effectively. This language-enhanced coordination becomes critical in complex scenarios where agents must handle ambiguous or evolving tasks that require continual communication and mutual understanding. The exploration of these capabilities opens up new possibilities for designing more intelligent and flexible multi-agent systems capable of operating in unpredictable, real-world environments.



Guo et al.~\cite{guo2024large} reviewed LLM-based multi-agent frameworks, but the emphasis of that paper was not on MARL. Unlike their paper, this letter focuses more on the MAS that tries to accomplish a task cooperatively. In addition to that, there are several surveys on the topic of MARL~\cite{nguyen2020deep, hernandez2019survey, gronauer2022multi} and single agent LLM-based RL~\cite{luketina2019survey, cao2024survey}, but none of them is dedicated to LLM-based MARL. Therefore, \emph{we claim that we are among the first to provide a systematic overview of the LLM-based MARL problem and provide potential future research directions.}

The remainder of this letter is organized as follows. We first introduce the problem of MARL and provide a brief overview of conventional, i.e., non-LLM-based, MARL, and single-agent LLM-based RL, in Sect.~\ref{sect:preliminary}. Then, we will survey the existing LLM-based MARL frameworks in Sect.~\ref{sect:survey}. After that, we will discuss the challenges and future research directions for this field in Sect.~\ref{sect:open_research}. Finally, we will conclude the letter in Sect.~\ref{sect:conc}.


\section{Preliminaries} \label{sect:preliminary}
In this section, we will first introduce the problem of MARL (Sect.~\ref{sect:pre:marl}). Then, we will briefly discuss conventional non-LLM-based MARL in Sect.~\ref{sect:pre:trad_marl}. To prepare the ground for LLM-based MARL, we will introduce LLM-based single-agent RL in Sect.~\ref{sect:pre:srl}.

\subsection{MARL Problem Definition} \label{sect:pre:marl}
MARL can be modeled with the Decentralized Partially Observable Markov Decision Process~(Dec-POMDP)~\cite{oliehoek2016concise}, an extension to a multi-agent manner of the Markov Decision Process~(MDP).
An MDP for $N$ agents consists of a set of states $\mathbf{s} \in \mathcal{S}$, which describes all the configurations for the participating agents, a set of actions $\mathcal{A}_1, ..., \mathcal{A}_N$ and a set of observations $\mathcal{O}_1, ..., \mathcal{O}_N$. Each agent $i$ has a policy $\pol_i: \mathcal{O}_i \times \mathcal{A}_i \mapsto [0, 1]$ parameterized by $\theta_i$. We denote deterministic policies by $\cpol_i: \mathcal{O}_i \mapsto \mathcal{A}_i$. The environment will generate the next state based on the state transition function $\mathcal{T}: \mathcal{S} \times \mathcal{A}_1 \times ... \times \mathcal{A}_N \mapsto \mathcal{S}$. Each agent will receive a reward from the environment as a function of state and action $r_i: \mathcal{S} \times \mathcal{A}_i \mapsto \mathbb{R}$ as well as an individual observation that is correlated with the state, $o_i:\mathcal{S} \mapsto \mathcal{O}_i$. Each agent tries to maximize its total expected return $R_i = \sum_{t=0}^T \gamma^t r_i^t$, where $\gamma$ is a discount factor, and $T$ is the total time length. A key difference between Dec-POMDP and normal MDP is the partial observability, i.e., for one agent, the actions of other agents and the subsequent outcomes are not directly observable, thereby increasing the difficulty of solving the problem. Due to this partial observability, individual uncoordinated learning frameworks will not work well. Typical deep MARL frameworks adopt the actor-critic structure, where actors are trained to output the action given the observation, and the critics output a score to judge whether these actions are good in the long-term horizon.



\subsection{Traditional MARL} \label{sect:pre:trad_marl}
To solve the problem of Dec-POMDP, many frameworks have been proposed.
These frameworks can be roughly categorized into two classes: learning-to-cooperate and learning-to-communicate.

\textbf{Learning to coordinate:}
The first kind of approach, such as QMIX~\cite{rashid2020monotonic}, QTRAN~\cite{son2019qtran}, MADDPG~\cite{lowe2017multi}, MAPPO~\cite{yu2022surprising}, and many others~\cite{sunehag2018value, rashid2020weighted, wang2021qplex, ackermann2019reducing, wang2020dop, zhang2021fop}, assumes that through centralized training with ideal communication, agents can learn to work with each other during the centralized training; therefore, communication is not needed during execution. In other words, these approaches expect the agents to learn to adapt to other agents' behavior patterns.
These approaches can also be classified as policy-based and value-based approaches. Policy-based approaches typically adopt the actor-critic architecture where actors are trained to make decisions, and critics approximate the long-term return and provide feedback to the actors.
Value-based approaches learn optimized joint Q values given the team's observations and actions. A problem that often happens in this situation is the credit assignment problem, where the critic needs to determine the contribution of each agent to the performance.

\textbf{Learning to communicate:}
In communication-based approaches, agents are equipped with the capability to share information through various means, such as adjusting the content of the shared messages~\cite{foerster2016learning} or optimizing the structure of the communication network~\cite{das2019tarmac}. This explicit inter-agent communication facilitates coordinated strategies and is crucial in dynamic environments where conditions and objectives may frequently change~\cite{sukhbaatar2016learning, hoshen2017vain}. Effective communication enables agents to form coalitions to achieve common goals, adapt to peers' actions, and optimize collective outcomes, improving system performance in tasks ranging from cooperative manipulation to competitive strategic games~\cite{foerster2016learning}. Protocols for communication, often learned during training, leverage advanced techniques such as differentiable interagent learning algorithms, which refine communication patterns based on environmental feedback~\cite{jiang2018learning, mordatch2018emergence, shen2021graphcomm}. In addition, frameworks for learning emergent communication protocols/languages have also been proposed~\cite{gupta2020networked, lazaridou2020emergent}. These frameworks encourage the agents to learn a certain ``language" that is understandable by other agents and encodes certain information.

\subsection{LLM-based Single-Agent RL} \label{sect:pre:srl}

As LLMs demonstrated their abilities in various tasks, several LLM-based decision-making frameworks have been proposed.
These frameworks are not necessarily RL frameworks because many of them are open-loop, meaning that the feedback/reward from the environment is not used during the decision-making process. Instead, many frameworks simply leverage the generalizability of LLMs and the general knowledge they contain to solve problems. Typically, in these works, a few examples of how the LLMs are expected to solve the problem are provided, and the LLMs can generalize from these examples to new problems.

\textbf{Open-loop LLM-based RL:}
Among these frameworks, we will summarize some significant contributions. Yao et al.~\cite{yao2023react} proposed ReAct, in which the LLM is prompted to generate ``thoughts" to solve the problem given the observation, allowing the model to dynamically adjust and refine its strategies in response to changing environmental cues and task demands. Based on ReAct, Shinn et al.~\cite{shinn2024reflexion} proposed Reflexion, which uses a few-shot verbal feedback to enhance decision-making capabilities. Reflexion processes feedback from interactions within task environments into textual summaries, which are then used to augment the model's episodic memory.
Prasad et al.~\cite{prasad2023adapt} proposed ADaPT, where LLMs learn to decompose the task into subtasks through short examples.
Although these approaches can achieve decent performances in reasoning or word-based games, they are constrained by the knowledge the LLMs have and could be biased for certain problems. More importantly, the reward, one of the most important signals from the environment, is not considered.

\textbf{Closed-loop LLM-based RL:}
There are also LLM-based RL frameworks that incorporate feedback for closed-loop control.
Paul et al.~\cite{paul2023refiner} proposed Refiner, in which a fine-tuned LLM is used to provide feedback on policy decisions. Zhang et al.~\cite{zhangsimple} introduced a framework that uses feedback from LLMs to enhance credit assignment in RL tasks. Their work targeted sparse reward environments and leveraged the rich domain knowledge available in LLMs to dynamically generate and refine reward functions. To improve sample efficiency, the authors proposed sequential, tree-based, and moving target feedback, facilitating more targeted exploration and reducing redundancy in state exploration. Yao et al.~\cite{yao2024retroformer} proposed Retroformer, where a frozen LLM is used as the policy, while another smaller LM is trained to provide verbal feedback on the decisions based on the reward. Murthy et al.~\cite{murthy2023rex} proposed REX, adopting the Monte-Carlo Tree Search~(MCTS) algorithm as the basis to solve problems. The Upper Confidence Bound~(UCB) technique is adopted to guide the agent's exploration.

Besides the aforementioned work that uses LLMs as RL policies, multi-modal LLMs that are trained on RL tasks such as robot control (e.g., PaLM-E~\cite{driess2023palm}) and models for grounding languages to actions~\cite{huang2022language, brohan2023can} have also been proposed. These models can achieve decent zero-shot performances in several robotic tasks because of their parameter scale.

\section{Existing LLM-based MARL} \label{sect:survey}

\begin{table*}[!t]
\centering
\caption{Existing LLM for MARL frameworks with an emphasis on multi-agent coordination.}
\label{tab:my-table}
\begin{tabular}{c|clcl}
\hline
\textbf{Framework} & \textbf{Application} & \multicolumn{1}{c}{\textbf{Dataset/Simulator}} & \textbf{Training} & \multicolumn{1}{c}{\textbf{LLM Role}} \\ \hline
DyLAN~\cite{liu2023dynamic} & Reasoning, Coding & \href{https://github.com/hendrycks/math/}{MATH}, \href{https://github.com/hendrycks/test}{MMLU}~\cite{hendrycksmath2021, hendrycks2021ethics}; \href{https://github.com/openai/human-eval}{HumanEval}~\cite{chen2021codex} & \xmark & Decision, Communication \\
FAMA~\cite{slumbers2023leveraging} & Text Game, Driving & \href{https://github.com/flowersteam/Grounding_LLMs_with_online_RL/tree/main/babyai-text}{BabyAI-Text}, Traffic Junction~\cite{sukhbaatar2016learning} & \cmark & Decision, Communication \\
Chen et al.~\cite{chen2023multi} & Consensus Seeking & \href{https://github.com/WestlakeIntelligentRobotics/ConsensusLLM-code/releases/tag/v1.0.1}{Generated Data} & \xmark & Decision \\
Li et al.~\cite{li2023theory} & Path Planning & Close-source simulator & \xmark & Decision, Communication, Theory of Mind \\
CoELA~\cite{zhang2024building} & Multi-Agent Planning & \href{https://github.com/threedworld-mit/tdw}{TDW-MAT}, \href{https://github.com/xavierpuigf/watch_and_help}{C-WAH}~\cite{puig2020watch} & \cmark & Decision, Communication, Memory \\
SMART-LLM~\cite{kannan2023smart} & Multi-Agent Planning & \href{https://github.com/SMARTlab-Purdue/SMART-LLM/tree/master/data}{Proposed Benchmark Dataset} & \xmark & Decision, Planning \\
RoCo~\cite{mandi2023roco} & Motion Planning & \href{https://github.com/MandiZhao/robot-collab/tree/main/rocobench}{RoCoBench} & \xmark & Decision, Planning \\
Co-NavGPT~\cite{yu2023co} & Semantic Navigation & \href{https://aihabitat.org/datasets/hm3d/}{Habitat-Matterport 3D}~\cite{ramakrishnan2021hm3d} & \xmark & Planning \\
Guo et al.~\cite{guo2024embodied} & Multi-Agent Cooperation & \href{http://virtual-home.org/}{VirtualHome-Social} & \xmark & Decision, Communication \\
MetaGPT~\cite{hong2023metagpt} & Coding & \href{https://github.com/openai/human-eval}{HumanEval}~\cite{chen2021codex}, \href{https://github.com/google-research/google-research/tree/master/mbpp}{MBPP}~\cite{austin2021program} & \xmark & Code Generation, Communication\\ \hline
\end{tabular}
\end{table*}

Although LLM-based MARL frameworks have not been widely studied, there is still some work focused on this topic.

\textbf{MARL for problem solving:}
Huang et al.~\cite{huang2024far} introduced $\gamma$-Bench, which encompasses a variety of multi-agent games to assess these models. Their work included a detailed analysis of different versions of the GPT models, which demonstrated a systematic improvement in their game ability. This framework demonstrated the enhanced performance of newer LLM versions, such as GPT-4, and the potential to augment these models with reasoning techniques such as CoT. Liu et al.~\cite{liu2023dynamic} proposed Dynamic LLM-Agent Network~(DyLAN), a framework that studied the capabilities of LLM-agent collaborations for complex reasoning and code generation tasks. Unlike previous methods that used static architectures, DyLAN dynamically adjusted agent interactions based on real-time performance and task demands, incorporating features such as inference-time agent selection and an early stopping mechanism. This allowed DyLAN to enhance computational efficiency and optimize the contribution of individual agents through an unsupervised scoring metric, the agent importance score.
Slumbers et al.~\cite{slumbers2023leveraging} introduced the Functionally-Aligned Multi-Agents~(FAMA) framework by integrating a
centralized critic architecture and allowing natural language communication between agents. The framework aligns LLMs to the functional needs of the environment through an online fine-tuning process, which adjusts the LLM's pre-trained knowledge to better fit the specific task requirements. Additionally, FAMA allows for intuitive inter-agent communication in natural language, making the coordination more efficient and human-interpretable. Chen et al.~\cite{chen2023multi} present a study on the dynamics of consensus seeking in multi-agent systems driven by LLMs. The authors focused on the inter-agent negotiation processes, where each agent starts with a unique numerical state and negotiates to reach a unified consensus. They also provided insights on how different factors, such as agent personality (stubborn vs. suggestible), agent number, and network topology, influence the negotiation and consensus process. Li et al.~\cite{li2023theory} explored Theory of Mind~(ToM) modeling with LLMs generating communication messages and beliefs about the environment and other agents. Hong et al.~\cite{hong2023metagpt} proposed MetaGPT, where agents share messages with all other agents in a message pool and agents can subscribe to messages related to their task.

\textbf{MARL for embodied applications:}
Other than the aforementioned MARL frameworks for problem solving, there are also LLM-based MARL frameworks for embodied application. Zhang et al.~\cite{zhang2024building} proposed a Cooperative Embodied Language Agent~(CoELA), a modular framework that integrates LLM to improve communication and collaborative decision-making among multiple agents. The modular structure includes a perception module for interpreting sensory data, a memory module for retaining and recalling environmental and task-related information, a communication module to facilitate inter-agent dialogue, a planning module for strategic decision making, and an execution module for carrying out planned actions. By incorporating LLMs into the memory, communication, and planning modules, the framework enables agents to utilize natural language to improve both understanding and execution of cooperative tasks.
Kannan et al.~\cite{kannan2023smart} introduced SMART-LLM, a framework that integrated LLM with multi-agent robot task planning to translate high-level instructions into executable strategies for robot teams. By structuring task planning into sequential phases of decomposition, coalition formation, and allocation, SMART-LLM generates robot actions to achieve complex objectives. Their approach leveraged the cognitive processing power of LLMs to enhance the comprehension and execution capabilities of robot systems.
Mandi et al.~\cite{mandi2023roco} introduced RoCo, a multi-robot arm collaboration framework with each arm equipped with an LLM agent. The LLM agents are responsible for coordination among agents by communicating with other LLM agents and path planning.
Yu et al.~\cite{yu2023co} introduced Co-NavGPT, an LLM-based multi-agent navigation framework. However, unlike other frameworks where multiple LLMs are employed, in Co-NavGPT, only one LLM is used to assign frontiers to agents globally.
Guo et al.~\cite{guo2024embodied} studied the collaboration of multiple LLM-based agents on various tasks with a focus on communication and coordination among multiple agents. They proposed the Criticize-Reflect method with an LLM critic and an LLM coordinator. Table~\ref{tab:my-table} provides more details on these works.

In addition to LLM-based MARL, several works explored multi-agent interaction~\cite{wu2023autogen, park2023generative, li2024camel}, e.g., multi-agent conversation and gaming. However, these works 
fall out of the MARL scope;
we will not use too much space on them.


Overall, these studies illustrated that while the exploration into language-conditioned MARL is still nascent, it holds considerable promise for advancing the capabilities of MAS. Using natural language, these systems can achieve higher levels of coordination and understanding, which is essential for complex environments.

\section{Open Research Problems} \label{sect:open_research}
\begin{figure*}[htp]
\centering
\begin{tabular}{@{}c@{}}
\subfloat{\includegraphics[width=0.39\linewidth]{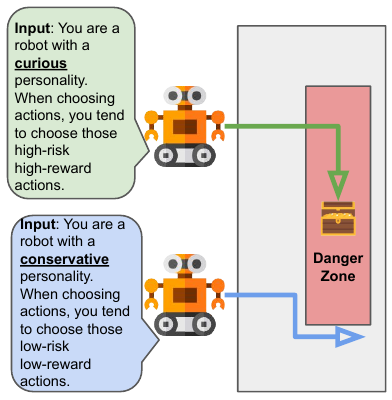}}\\ (a)
\end{tabular}\qquad 
\begin{tabular}{@{}c@{}}
\subfloat{\includegraphics[width=0.56\linewidth]{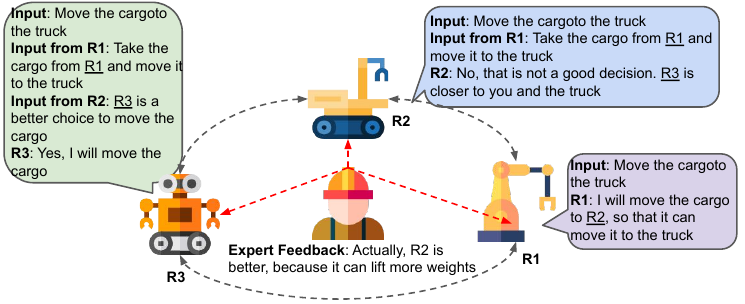}}\\ (b)
\\
\subfloat{\includegraphics[width=0.55\linewidth]{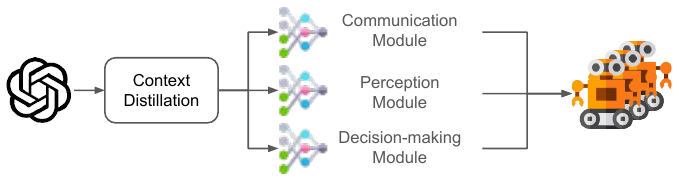}}\\ (c)
\end{tabular}
\caption{\label{fig:ideas}Potential research directions for language-conditioned Multi-Agent Reinforcement Learning~(MARL). (a)~Personality-enabled cooperation, where different robots have different personalities defined by the commands. (b)~Language-enabled human-on-the-loop frameworks, where humans supervise robots and provide feedback. (c)~Traditional co-design of MARL and LLM, where knowledge about different aspects of LLM is distilled into smaller models that can be executed on board.}
\vspace{-0.1in}
\end{figure*}

Despite the research efforts mentioned above, language-conditioned MARL is still an unexplored field with many unexplored aspects. To inspire more research in this field, we provide several research directions in this section. Specifically, we discuss four potential research directions: i)~\emph{personality-enabled cooperation} (Sect.~\ref{sect:open:person}), ii)~\emph{language-enabled human-in/on-the-loop frameworks} (Sect.~\ref{sect:open:human}), iii)~\emph{traditional MARL and LLM co-design} (Sect.~\ref{sect:open:codesign}), and iv)~\emph{safety and security in MAS} (Sect.~\ref{sect:open:safe}). Fig.~\ref{fig:ideas} also provides a more vivid demonstration of these research ideas.

\subsection{Personality-enabled Cooperation} \label{sect:open:person}

Previous work~\cite{szot2023adaptive, chen2023multi} has shown that different personalities in MARL frameworks can produce promising results. This idea can be naturally extended to language-conditioned MARL frameworks.
In these frameworks, agents are distinguished by their assigned personalities. For example, an agent with a ``curious" personality will tend to explore the environment, while an agent with a ``conservative" personality will tend to stay in the safe areas. A team of agents with a combination of different personalities can often achieve better performance than those with the same personality.
In traditional MARL frameworks, these personalities are encoded in the agents' model parameters, i.e., the weights of their models. However, with LLMs as agents, personalities can be assigned to agents by prompts, in which narratives about the agent's personality will be provided. 

Another potential advantage of language-conditioned MARL with personalized agents is the ability to handle conflicts and negotiate solutions more effectively. Agents can be trained to understand and generate language-based responses that consider the perspectives and goals of other agents, facilitating a negotiation process that mirrors human interaction. This capability is particularly useful in scenarios where agents must share resources or decide on joint actions that impact the collective outcome.

However, implementing these personalized language behaviors in agents presents several challenges. The primary concern is ensuring that language models do not perpetuate or amplify undesirable biases that could lead to unfair or inefficient outcomes. Additionally, the complexity of training such models increases as they must not only understand and generate appropriate responses, but also adapt their linguistic style based on the evolving context of the interaction.

Future research could focus on developing frameworks that can effectively integrate personality-driven language models into MARL systems. This integration involves creating robust prompts with memories that encode the information from past experiences in a wide range of interactive scenarios, allowing agents to learn from both their successes and failures. Furthermore, evaluating these systems will require new metrics that can assess not just the efficacy of task performance but also the appropriateness and effectiveness of communication between agents.

Another direction of research is to explore competitive agents instead of cooperative agents. However, the competition here should be benign, which means that the agents compete to achieve the same goal. 
By addressing these challenges, language-conditioned MARL with diverse agent personalities has the potential to advance the field of artificial intelligence.

\subsection{Language-enabled Human-in/on-the-Loop Frameworks} \label{sect:open:human}

One of the direct advantages of language-conditioned MARL frameworks is the possibility of involving humans in or on the loop. To illustrate, human-in-the-loop frameworks~\cite{abel2017agent, liang2017human, luo2023human} involve humans as agents that can generate actions to affect the environment, while human-on-the-loop frameworks~\cite{christiano2017deep} regard humans as supervisors without directly being involved in the decision-making process. 

In human-in-the-loop setups, humans actively participate in the learning process, often providing corrective feedback or rewards to shape agent behaviors in real time. This direct interaction helps in refining the agent's actions and strategies, making them more aligned with human-like reasoning and ethical standards. For example, a human could guide an agent away from potential pitfalls in its learning process that might not be immediately apparent through algorithmic reinforcement signals alone.
On the other hand, human-on-the-loop frameworks play a crucial oversight role. Here, humans monitor the system's performance and intervene only when necessary. This approach is particularly valuable in applications where autonomous operations are preferable, but human oversight is necessary to ensure safety and compliance with regulatory standards. For example, in autonomous driving, while the system can handle most driving tasks, a human supervisor may only need to intervene in complex or hazardous road conditions, ensuring that the system operates within safe limits without requiring constant human control.

Both of these human roles within language-conditioned MARL can benefit significantly from the integration of natural language. Language serves as a versatile interface that enables clearer and more intuitive communication between humans and agents. Agents can report their status, explain their decisions, or even ask for clarification in human-understandable language, improving the effectiveness of human interventions.
Furthermore, the use of language can facilitate the transfer of knowledge between agents by allowing them to share insights or strategies in a comprehensible format. In scenarios involving multiple agents with varying roles, language can help maintain coherence and unity of purpose across the team, guiding less experienced agents through complex tasks or strategies articulated by more experienced ones or even by human supervisors.

Future research could explore optimizing these interactions between human supervisors and agents, possibly by developing advanced language models that can understand and generate more context-aware, situation-specific dialogue. Furthermore, ensuring that language-based communications are not only informative, but also prompt and actionable will be crucial for the practical deployment of such systems in real-world applications. This balance between automation and human oversight, facilitated by natural language, promises to enhance the robustness and reliability of multi-agent systems, pushing the boundaries of what automated systems can achieve while ensuring they operate under safe and ethical guidelines.


\subsection{Traditional MARL and LLM Co-Design} \label{sect:open:codesign}
Since LLMs tend to have large sizes, especially those pre-trained models, performing inference on-board on robot hardware is not practical. A popular way towards resource-efficient computing is through Parameter-Efficient Fine-Tuning~(PEFT) techniques~\cite{hu2021lora, xin2024mmap, xin2024vmt, xin2024parameter} combined with quantization. However, this kind of approach still requires inference through the large LLM network, which is impractical for small robots. To make this happen, we envision a co-design framework of traditional MARL policies and the LM models. A typical design for such systems could be to use the LLM model as a centralized critic to guide the training of the actors. This design follows the CTDE scheme introduced in Sect.~\ref{sect:pre:trad_marl}, where the critic will be removed during execution. To leverage communication during execution, we can distill the knowledge from the LLMs about communication into smaller models that can be executed onboard.

One potential development is the refinement of the distillation process, which aims to transfer knowledge from LLMs to more compact models suitable for deployment on less powerful hardware, such as robots or Internet of Things~(IoT) devices.
A promising direction in this direction would be in-context distillation~\cite{huang2022context, snell2022learning}, where the teacher model is an LLM with a pre-defined context. For example, for controlling warehouse robots, the context can be refined to tell the LLM to avoid people and collisions. By focusing on the essential features necessary for the communication and decision-making learned by the LLM, smaller models can execute complex tasks effectively with a fraction of the computational overhead. In addition, to facilitate effective communication between agents during execution, specialized communication protocols could be designed. These protocols would utilize the distilled models to ensure that critical information, as understood and processed by the LLM during the training phase, is efficiently conveyed between agents. This approach not only conserves bandwidth, but also optimizes the real-time decision-making process, allowing for dynamic adjustments based on the operational environment and agent states.

Additionally, the co-design framework can be enhanced by integrating adaptive mechanisms that allow the MARL system to recalibrate its strategies based on feedback from the operational environment. Such adaptive systems could dynamically adjust the compression level of the distilled models or modify the communication protocols based on the complexity of the tasks and the computational capabilities available at that time. This flexibility would be particularly useful in environments where conditions change rapidly or unpredictably, requiring swift responses from the agent collective. Furthermore, the implementation of this co-design framework would benefit significantly from the development of specialized hardware tailored to the execution of compressed models. This hardware could optimize the execution of neural network operations, potentially in a power-efficient manner, which is critical for mobile or embedded systems.

\subsection{Safety and Security in MAS} \label{sect:open:safe}

Ensuring the safety and security of MAS is critical, especially as these systems are increasingly deployed in diverse and potentially high-stakes environments. The integration of language models into MARL introduces unique challenges and vulnerabilities, from the manipulation of agent communication to the exploitation of model biases.

Many robotic operations have continuous action spaces, where the output of each agent's policy is a set of continuous values. Unlike discrete action spaces, which can be reformulated as multi-choice problems and solved by prompting the multi-choice question to the LLM, continuous action space is more tricky, especially in high-stake environments, for example, operation robots. Existing methods replace the last few layers of the LLMs with new layers that map the observation in languages to continuous action spaces. However, this kind of approach requires training the new layers in the desired environment, which might be inaccessible. Therefore, exploring alternative methods for integrating LLMs into the control loop of robots operating in continuous action spaces without the need for substantial retraining or modification of the LLMs is promising.

In addition to safety in actions, safety and security against potential attacks are also crucial in MAS. One way towards safety is through proactive measures. This includes the development of secure communication protocols between agents to prevent eavesdropping or the injection of malicious data that could lead to compromised decision-making. Communications encryption can be a fundamental aspect of this, ensuring that even if data transmissions are intercepted, the information remains protected. In addition, securing the language model training process against adversarial attacks is crucial. Adversarial training, which involves exposing the system to a wide range of attack vectors during the training phase, can help models learn to resist or mitigate these attacks in deployment. In addition, input validation techniques can be employed to filter out potentially harmful or misleading inputs that could cause the system to behave unpredictably. This is particularly important in scenarios where agents interact with humans or systems outside the controlled environment and are exposed to a broader range of language inputs and behaviors.

Despite the best proactive defenses, systems may still encounter unforeseen vulnerabilities post-deployment. Thus, reactive strategies are necessary to quickly address any breaches or failures. This can involve real-time monitoring of agent behaviors and communications to detect anomalies that may indicate a security breach or a failure in safety protocols. Once an anomaly is detected, the systems should be able to isolate affected agents and roll back their states to secure configurations.




\balance

\section{Conclusion} \label{sect:conc}
In this letter, we provide a brief overview of Multi-Agent Reinforcement Learning (MARL) based on conventional non-Large Language Model~(LLM)-based Multi-Agent Reinforcement Learning~(MARL), LLM-based single-agent RL, and existing LLM-based MARL frameworks.
These works paved the way for new ideas that we discuss in later sections. Specifically, we discussed potential research directions ranging from multi-agent personality to safety and security in the LLM-based Multi-Agent System~(MAS).
Although works are studying LLM-based MARL, the field is still to be explored and has significant potential because of the great ability of LLMs and their in-context and interpretable nature. With LLMs, designing MARL frameworks becomes more analogous to modeling the group learning process of animals or even humans, where knowledge is transferred or exchanged via natural languages. We hope, with this letter, that more research works can be enlightened and the boundary of multi-agent intelligence could be pushed further.

\bibliographystyle{ieeetr}
\bibliography{main}
\end{document}